# piSAAC: Extended notion of SAAC feature selection novel method for discrimination of Enzymes model using different machine learning algorithm

Zaheer Ullah Khan[1], Dechang Pi[*1], Izhar Ahmed Khan[1], Asif Nawaz[1], Jamil Ahmad[1], Mushtaq Hussain[2]

[1]School of Computer Science and Technology, Nanjing University of Aeronautics and Astronautics, China
[2]Department of Information Technology, The University of Haripur, Paksitan

Author's email addresses:
* Corresponding author : zaheerkhan@nuaa.edu.cn

**Abstract**

**Enzymes and proteins are live driven biochemicals, which has a dramatic impact over the environment, in which it is active. So, therefore, it is highly looked-for to build such a robust and highly accurate automatic and computational model to accurately predict enzymes nature. In this study, a novel split amino acid composition model named piSAAC is proposed. In this model, protein sequence is discretized in equal and balanced terminus to fully evaluate the intrinsic correlation properties of the sequence. Several state-of-the-art algorithms have been employed to evaluate the proposed model. A 10-folds cross-validation evaluation is used for finding out the authenticity and robust-ness of the model using different statistical measures e.g. Accuracy, sensitivity, specificity, F-measure and area un-der ROC curve. The experimental results show that, probabilistic neural network algorithm with piSAAC feature extraction yields an accuracy of 98.01%, sensitivity of 97.12%, specificity of 95.87%, f-measure of 0.9812and AUC 0.95812, over dataset S1, accuracy of 97.85%, sensitivity of 97.54%, specificity of 96.24%, f-measure of 0.9774 and AUC 0.9803 over dataset S2. Evident from these excellent empirical results, the proposed model would be a very useful tool for academic research and drug designing related application areas.**

## I. INTRODUCTION

Enzymes are proteins in nature and act as a biological catalyst(Mohamad, Marzuki, Buang, Huyop, & Wahab, 2015; Zhao et al., 2016). It speeds up the rate of the different biochemical and metabolic reaction in the living organism(Jullesson, David, Pfleger, & Nielsen, 2015). The enzyme can work within a narrow range of temperature and pH(Khan, Hayat, & Khan, 2015) The pH value greatly affect the enzyme activity. It has also been observed experimentally, that at optimum pH value in a chemical environment reflect strong enzyme activity effectiveness(Cummings, Murata, Koepsel, & Russell, 2014; Y. Zhang, Ge, & Liu, 2015). Role of Enzymes are very important in drug industries(Choi, Han, & Kim, 2015), cell growth(Li, Liao, Zhang, Du, & Chen, 2011) and cell communication(Alberts et al., 2002; Bu & Callaway, 2011). The Enzyme catalytic nature is due to its contemplative specificity, selectivity and catalytic efficiency(Fried & Boxer, 2017). pH and temperature have a crucial effect on the enzyme efficiency(Nakhil Nair, Carl, & ZhaoH.; Sarethy et al., 2011). Most enzymes endure their optimum reactivity in the scale of 6 and 8(Kemper Talley, 2010; Zhang Li, Liang Fan, & Chun Zuo, 2013). There is enough rich literature on enzyme stability in the field of biophysical and biotechnological(Durán, Silveira, & Durán, 2015). The stability of the enzyme is very essential in any particular biochemical reaction environment, in which it is active. In such scenario, pH is one major bottleneck, in extending its implication and their applications (Geierstanger, Jamin, Volkman, & Baldwin, 1998; Min, 2014). For that purpose, some researchers have developed many sequences based computational tools in the field of proteomics. Many of theoretical discrete methods have achieved considerable success on the basis of primary and secondary protein sequence structure(Wei Chen, Ding, Feng, Lin, & Chou, 2016; Jianhua Jia, Liu, Xiao, Liu, & Chou, 2016). These amino acid sequences adhere to a strong correlation to the external environment of the enzyme (Dubnovitsky, Kapetaniou, & Papageorgiou, 2005; Kelch et al., 2007). There are many research studies, which shows some discrete method to formulate the primary sequence information. Similarly (G. Zhang, Li, & Fang) has proposed a computation model for discrimination of acidic and alkaline enzyme. They used random forest algorithm using secondary structure amino acid composition. Likewise, (Zhang Li et al., 2013) also proposed a method for discrimination of enzymes. Then (Chou) has suggested a comprehensive review and methods for evaluating the amino acid primary sequence along with their order information and correlation factor, and published a series of research papers. Onward (Carugo, 2008; M Hayat, A Khan, & M Yeasin; Khan et al., 2015) has also added comprehensive literature over discrimination of enzymes. Although there is a lot of research contribution to this topic by pattern recognition and machine learning. We have extended our proposed feature extraction methodology in the notion of prior SAAC (split amino acid composition). Several state-of-the-arts algorithms

were used to study and yield the optimum prediction of acidic and alkaline enzyme over the proposed model

## II. MATERIAL AND METHODS

Benchmark Dataset

Selection of benchmarked dataset is the first choice for building a promising computation model (Khan et al.; Zhang Li et al.; ZU Khan) The dataset was obtained from (H Lin, W Chen, & H Ding). Which is composed of 54 acidic and 68 alkaline enzymes, constituting a total of 122 enzymes as dataset1. The underlined dataset refined over criteria of having optimal pH 5.0 for alkaline and 9.0 for acidic. The selection was also made subject to 25% CDHit for having more than 25% identical sequence existence. Another dataset comprehended by(Fan, Li, & Zuo, 2013) has been selected by having 105 acidic and 112 alkaline enzymes represented as dataset2.

### A. Proposed Model framework

A diagrammatic presentation helps in presenting a model in a clear logic and useful intuitiveness insight for understanding its complicated relationship. The proposed model framework is given in Figure 1. Many existing approaches have used meta or simple approaches for feature generation, which were unable to consume to the true discriminative feature information. To fill that niche gap, we have presented an invative approach by yielding a extended feature out of Split Amino Acid features. A rigorous feature elimination method (RFE with Fold Cross-validation) was used to remove noisy, and vague features. The optimum feature greatly enhance the classification power of the baseline classification algorithm.

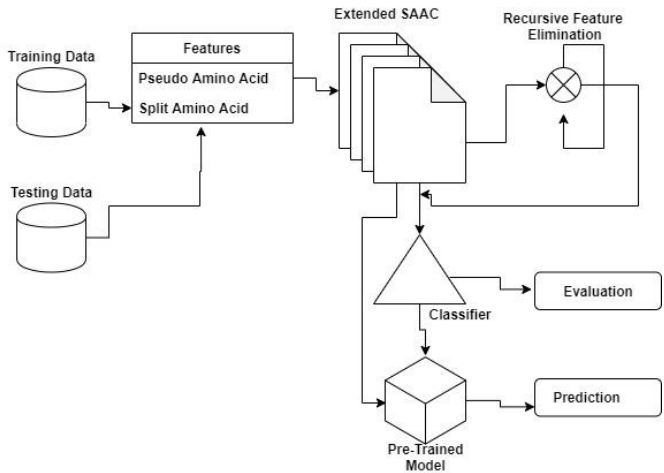

Fig.1. Proposed Model schematic Flow diagram

### B. Feature Extraction approaches

One of the major bottlenecks of the machine learning-based algorithms is that it only processes discrete vectors rather than a raw biological sequence. Therefore, several feature extraction techniques have been developed to transform a raw biological sequence into a discrete feature vector.

**Pseudo Amino Acid Composition**

Simple amino acid composition only represents the core native 20 amino acid cumulative occurrences frequency in the amino acid strait, resulting in a 20D feature vector, represented in equation Eq.1.

$$P_i = \frac{n_i}{L} \quad (1)$$

In the equation Eq.1 the pi represents the cumulative frequency of each native protein and L is the total length of the protein sequence. But however, the study shows a different tendency of proteins role mechanism in the formulation of the protein secondary structure. Therefore it is evident that the locational information and ordering also a vital role in the exact discrimination of the enzymes(Mei & Zhao, 2018). Pseudo amino acid composition, incorporate the order information along with the correlation factor(Chou, 2001; J. Jia, Liu, Xiao, Liu, & Chou, 2015). Chao, concept of pseudo amino acid composition is widely used and adopted almost every field of proteomics. The pseudo amino acid composition can be described via

$$p = [p_1 \ldots p_{20}, p_{20+1}, \ldots p_{20+\lambda}]^T \quad (2)$$

Whereas T represent the transposition vector, and $p_1 \ldots p_{20}$ represents native amino acid occurrences frequency with additional parameter, that exhibits the correlation factor of amino acid subject to the hydrophobicity, hydrophilicity, charge polarity.

$$\tau 1 = 1L - 1i = 1L - 1J\,i\,, +1\tau 2 = 1L - 2i = 1$$
$$L - 2J\,i\,, +2\tau 3 = 1L - 3i = 1L - 3J\,i\,,i + 3$$
$$.\tau \lambda = 1L - \lambda\,i = 1L - \lambda J\,i\,, +\lambda \quad (3)$$

In equation (3) L represents the length of the protein sequence, whereas $\tau_1$ represents first correlation factor till last $\tau_\lambda$ rank correlation. We have selected best value of lambda λ=28 for first 28 order correlation factor for rank information. As we have also utilized hydrophobicity, hydrophilicity, charge and polarity, which consequently, makeup a feature vector of $20 \times (20 + 112) - D = 20 \times 132 - D$.

**Table depicting feature values of twenty native amino acids**

| AA | Hydrophobicity | Hydrophilicity | Rigidity | Flexibility | Irreplaceability |
|----|----|----|----|----|----|
| A | 0.62 | -0.50 | -1.33 | -3.10 | 0.52 |
| C | 0.29 | -1.00 | -1.51 | 0.95 | 1.12 |
| D | -0.90 | 3.00 | -0.20 | 0.42 | 0.77 |
| E | -0.74 | 3.00 | -0.36 | 2.00 | 0.76 |
| F | 1.19 | −2.50 | 2.87 | -0.46 | 0.86 |
| G | 0.48 | 0.00 | −1.09 | −2.74 | 0.56 |
| H | −0.04 | −0.50 | 2.26 | −0.22 | 0.94 |
| I | 1.38 | −1.80 | −1.74 | 0.42 | 0.65 |
| K | −1.50 | 3.00 | −1.82 | 3.95 | 0.81 |
| L | 1.06 | −1.80 | −1.74 | 0.42 | 0.58 |
| M | 0.64 | −1.30 | −1.74 | 2.48 | 1.25 |
| N | −0.78 | 0.20 | −0.20 | 0.42 | 0.79 |
| P | 0.12 | 0.00 | 1.97 | −2.40 | 0.61 |
| Q | −0.85 | 0.20 | −0.36 | 2.00 | 0.86 |
| R | 2.53 | 3.00 | 1.16 | 3.06 | 0.60 |
| S | −0.18 | 0.30 | −1.51 | 0.95 | 0.64 |
| T | −0.05 | −0.40 | −1.64 | −1.33 | 0.56 |
| V | 1.08 | −1.50 | −1.64 | −1.33 | 0.54 |
| W | 0.81 | −3.40 | 5.91 | −1.00 | 1.82 |
| Y | 0.26 | −2.30 | 2.71 | −0.67 | 0.98 |

**Pi-SAAC (split amino acid composition)**

It is evident from the research and experimentally drawn results, that the sequence ordering greatly affect, the role of a particular protein. Getting crucial and intrinsic features, concealed in segments of a primary sequence (M Hayat et al.; Khan et al.) remains of value in the design of any computational model. In order to extract intrinsic features, we have proposed a novel feature extraction method, in which primary protein sequence is divided into segments.

Where each terminus is subject to n/5 size interval. Where n is the number of the terminus. A given protein sequence is denoted by Sn

$$S_n = [R_1 R_2 R_3 ... R_L] \quad (4)$$

Whereas R1, R2 represent the first and second residue of the sequence Sn, and RL represents the total number of residues. To utilize and get the optimal correlated features, we have divided the sequence into subsequence. This interval makes equal distribution of features over a split of 5 parts. These intervals are made subject to the q in Eq.5.

$$m = q \times d + r \quad (5)$$

Since we know that, from elementary mathematics, Eq.5 m is the total length of the sequence, q is quotient, d is divisor and r is remainder.

$$S_n = S_1 \cup S_2 \cup S_3 \cup S_4 \cup S_5 = \bigcup_{k=1}^{5} S_k \quad (6)$$

In Eq.6 Sn represent the sum up feature vector sequence of five subsequences. Where these sub sequence intervals, this Sn represents the representative sequence of all subsequence amino acids. Mnemonically, the above Eq.3 sequence interval can be represented with Exclusive Left as EL, left, Middle, Right and Exclusive Right, which are shown as follow in equation Eq.4.

$$S_n = S_{EL(1,Q)} \cup S_{L(2Q,Q)} \cup S_{M(2Q,m-2Q)} \cup S_{R(Q,2Q)} \cup S_{ER(m-2Q,m-Q)} = \bigcup_{k=1}^{5} S_k \quad (8)$$

In the above Eq.6, Q represents the quotient, which can be derived from Eq.2 and m is the total number of sequence length. In Eq.5 we have designated different interval range for each subsequence interval, which distributes the biased feature vector into 5 equal sub sequence spaces. It's not only equally distributes the prime features over the entire span of vector but also exploit the prime biasedness of features over whole sequence vector. Equ.6 can be expressed as follow for each individual terminus.

$$f_u = [f_1 f_2 f_3 f_4 f_5 .... f_{20}]^T \quad (9)$$

Whereas $f_u = (u = 1,2,3,4,5 ... 20)$ are the normalized frequencies of 20 native amino acids in the sequence. A general pseudo amino acid composition vector Sn is calculated for each subsequent sub sequence over union making a resultant of 100*D vector.

### C. Recursive Feature Elimination

The outlier present in the data drastically degrades the performance of the computational model and often misleads classification prediction. The optimum selected feature prevents the model from the curse of dimensionality, avoids overfitting, reduces training time, and enhanced model generalizability. RFE is a wrapper type function selection algorithm. Technically, RFE is a shell-like resource selection algorithm that also uses internal resource selection based on filters. RFE works by finding a subset of resources, starting with all the resources in the training dataset, and successfully removing the resources until the desired number remains. This is achieved by adjusting a certain machine learning algorithm used in the core of the model, ranking resources by importance, eliminating less important resources, and re-adapting the model. This process is repeated until the specified number of functions remain.

### D. DNN(Deep Neural Network)

Over a decade deep neural network remains the dominant problem-solving underline architecture, which can learn any complex pattern and hidden information. Recently, deep learning has shown up state-of-the-art classifying [22][53], [68]–[75] model in term of accuracy and authenticity. A deep neural network layer can learn a high level of abstraction from a huge dimensional scattered data. It discovers an intricate underline structure of data using backpropagation algorithm by making the machine learn how to change its internal parameter for learning the representation from the previous layer weights. We have implemented famous python machine learning keras deep neural network library. We have used, 'relu' as activation function and 'sigmoid' as the prediction layer activation function. For optimization, we have used the stable and fast converging algorithm 'adam' with lr:0.001, beta_1:0.9, beta_2:0.9 with a decay of 0.00 and 'dropout' of 0.2 for avoiding model overfitting. Have employed sigmoid as a squash function.

### E. Performance Evaluation Metrics

To measure the quality of a model, two things are kept under consideration, i) Quantitative measure metrics and ii) cross-validation test. Different statistical model evaluation metrics were exercised to quantify the robustness and authenticity of the model[17-19][20-21]. Accuracy is a well-known statistical metrics used as a classifier correctness measure tool, sensitivity or recall measure true positive rate, specificity measure the true negative rate, MCC measures the model stability and ROC curve measure the overall model performance and authenticity in respect of just making random judging. F-measure is the harmonic mean of sensitivity and specificity operating in a range of 0 and 1. Cohens-kappa statistics, measure the inter-observer agreement, the reliability of the system. Average precision is a singular value metric to evaluate the model prediction.

$$Sn = 1 - \frac{N^\pm_+}{N^+} \qquad 0 \leq Sn \leq 1 \qquad (1)$$

$$Sp = 1 - \frac{N^\mp_-}{N^-} \qquad 0 \leq Sp \leq 1 \qquad (2)$$

$$Acc = 1 - \frac{N^\pm_+ + N^\mp_+}{N^+ + N^-} \qquad 0 \leq Acc \leq 1 \qquad (3)$$

$$MCC = \frac{1 - \left(\frac{N^\pm_+ + N^\mp_+}{N^+ + N^-}\right)}{\sqrt{\left(1 + \frac{N^-_+ - N^\pm_+}{N^+}\right)\left(1 + \frac{N^\pm_+ - N^\mp_+}{N^-}\right)}} \qquad -1 \leq MCC \leq 1 \qquad (4)$$

$$Fs = (2/recall^{-1} + precision^{-1}) \quad 0 \leq F1 \leq 1 \qquad (5)$$

$$kappa = 1 - \frac{1-p_0}{1-p_e} \qquad 0 \leq kappa \leq 1 \qquad (6)$$

In Eq. 1~6, Sn refers to sensitivity, Sp to specificity, Acc to accuracy and MCC to Mathew correlation coefficient, Fscore and Kappa statistics.

III. RESULT AND DISCUSSION

This section discussed the assessment of different classification algorithms trained over a different set of primitive In this proposed work, we have evaluated and analyzed the performance of several classification algorithms over SAAC feature extraction method. The performance measure of the proposed model using piSAAC is listed in Table.2 using dataset 1. There results were drawn by decision tree with accuracy of 80.53%, sensitivity of 80.11%, specificity 82.89%, F measure 0.8114 and AUC of 0.8212 on dataset1, 85.31% 84.43%, 85.01%, 0.8782, 0.8872 on dataset2 respectively. Comparatively random forest and Multi-Layer Perceptron with Linear kernel function 86.04%, 85.23%, 81.56%, 0.8243, 0.8665, and 85.81%, 84.76%, 86.00%, 0.8643, and 0.8532 as accuracy, sensitivity, specificity, f-measure and area under curve on dataset1 and 84.04%, 83.23%, 83.53%, 0.8343, 0.8762, 87.22%, 87.13%, 86.44%, 0.9011, 0.8813 on dataset 2, respectively. In this study we have employed a minimal SVM with least number of acting support vectors, which greatly improve the prediction accuracy of our proposed model, yielding 94.02%, 92.65%, 95.11%, 0.9316, and 0.9334 on dataset1, 93.88%, 94.90%, 92.21%, 0.9257, 0.9412 on dataset2 as Acc, sen, sp, f-measure and auc respectively. These results clearly indicate that the minimal svm perform greatly, by employing least number of support vectors on the comparatively light dataset. In contrast with the other classifiers, DNN outperform on dataset1 with highest prediction success rate compared with other classifier, resulting of 98.01%, 97.12%, 95.87%, 0.9812, 0.9500 on dataset1,

97.85%, 97.54%, 96.24%, 0.9774, 0.9803 on dataset2 as accuracy, sensitivity, specificity, f measure and AUC respectively. These empirical figures are shown in Table.2, Table.3 shows, in term of classification success rates of the proposed model. In contrast over all the datasets the proposed feature selection method, minimal SVM and DNN outperform respective to others classifiers. Respective Model Loss and Model train test Accuracy are given in Figure.2,3 and 4 respectivily.

**Table.2 Classifier success rate over *piSAAC* on dataset1**

| Classifier | Acc (%) | Sn (%) | Sp (%) | F Measure | AUC |
|---|---|---|---|---|---|
| DTC48 | 80.53 | 80.11 | 82.89 | 0.8114 | 0.8212 |
| MLP | 85.81 | 84.76 | 86.00 | 0.8643 | 0.8532 |
| KNN | 75.54 | 74.23 | 76.12 | 0.7617 | 0.7977 |
| SVM | 94.02 | 92.65 | 95.11 | 0.9316 | 0.9334 |
| NB | 85.02 | 88.00 | 82.87 | 0.8253 | 0.8834 |
| RF | 86.04 | 85.23 | 81.56 | 0.8243 | 0.8665 |
| SMO | 84.88 | 84.43 | 82.34 | 0.8266 | 0.8523 |
| Adaboost | 84.34 | 83.21 | 85.87 | 0.8423 | 0.8512 |
| DNN | 98.01 | 97.12 | 95.87 | 0.9812 | 0.9500 |

**Table.3 Classifier success rate over *piSAAC* on dataset2**

| Classifier | Acc(%) | Se(%) | Sp(%) | F-Measure | AUC |
|---|---|---|---|---|---|
| DTC48 | 85.31 | 84.43 | 85.01 | 0.8782 | 0.8872 |
| MLP | 87.22 | 87.13 | 86.44 | 0.9011 | 0.8813 |
| KNN | 84.34 | 88.21 | 77.21 | 0.8321 | 0.9122 |
| SVM | 93.88 | 94.90 | 92.21 | 0.9257 | 0.9412 |
| Naïve Bayes | 84.31 | 84.66 | 86.93 | 0.8452 | 0.9054 |
| Random Forest | 84.04 | 83.23 | 83.53 | 0.8343 | 0.8762 |
| SMO | 84.05 | 82.44 | 78.33 | 0.8143 | 0.8222 |
| Adaboost | 85.57 | 85.43 | 84.83 | 0.8321 | 0.8611 |
| DNN | 97.85 | 97.54 | 96.24 | 0.9774 | 0.9803 |

IV. COMPARISON ANALYSIS

Several other researchers have already worked on discrimination of acidic and alkaline enzymes and recombination spot using machine learning approaches. To compare our proposed approach results with existing methods can be viewed in the listing Table 4.

(Khan et al., 2015) represented most recently a discriminative computational model for discrimination of enzymes, with empirical figures on SAAC, 94.47%, 95.50%, 93.40%. 0.9300, as accuracy, sensitivity, specificity, f measure and AUC on dataset1, and 92.62%, 94.12%, 90.74%, 0.9900 on dataset2 respectively. Our proposed model is 4% and 6% respectively higher than those empirical results of 98.01%, 97.12%, and 95.87% and 0.9812, as an accuracy, sensitivity, specificity and AUC. Likewise, in contrast with (Fan et al., 2013) and (Hao Lin, Wei Chen, & Hui Ding, 2013) models results, as an accuracy, sensitivity, specificity and AUC were of 94.00%,

98.53%, 95.50%, 0.9600 and 94.40%, 94.60%, 94.30%, 0.9700 respectively, comparatively with our proposed model, of an accuracy of 98.01%, sensitivity 97.12%, specificity 95.87% and area under curve of 0.9812, which clearly state, that our proposed model is better and authentic based on these statistical figures.

**Table.4. comparision of proposed approch with exisiting methods**

| Methods | Acc (%) | Sn (%) | Sp (%) | AUC |
|---|---|---|---|---|
| (Fan et al., 2013) | 94.00 | 98.53 | 95.50 | 0.9600 |
| (Hao Lin, Wei Chen, & Hui Ding, 2013) | 94.40 | 94.60 | 94.30 | 0.9700 |
| (Khan et al., 2015) | 94.47 | 95.50 | 93.40 | 0.9300 |
|  | 92.62 | 94.12 | 90.74 | 0.9900 |
| (L. Zhang & Kong, 2018) | 83.72 | 75.51 | 90.52 | 0.0000 |
| Proposed | 98.01 | 97.12 | 95.87 | 0.9812 |

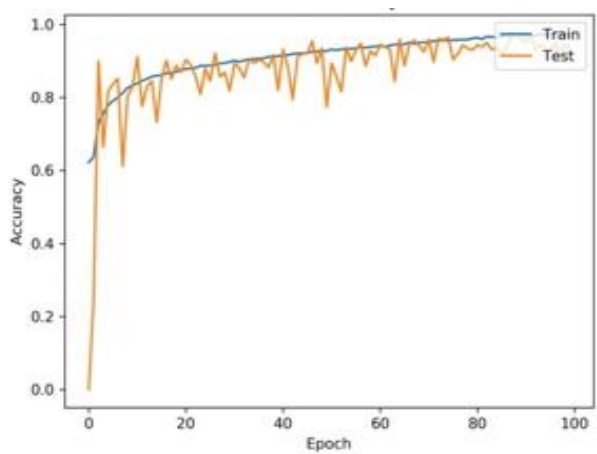

Figure 2: Model Accuracy Plot Training and Validation Test data

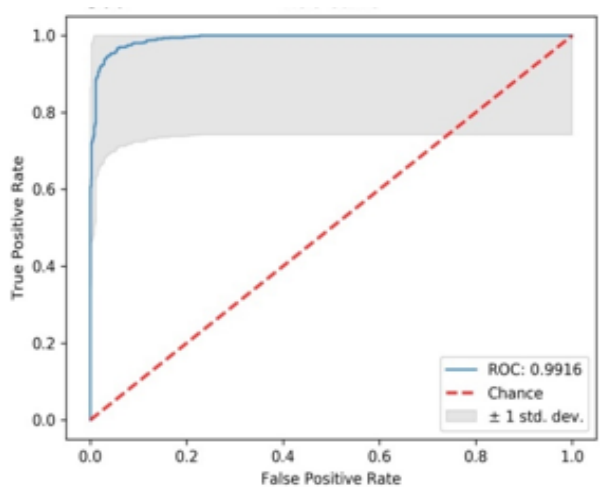

Figure.3: ROC Curve plotting of Proposed Model over Proposed Feature

## V. COMPARISON ANALYSIS

The most obvious objective of building an effective computational predictor is its generalizability. Table.4 lists detail comparison of the proposed model and the existing state of the art models, it can be seen that our model outperform the existing study by securing an obvious improvement in Acc, Sp, F-score and ROC curve.

Table.5. Proposed Model Comparison analysis with other studies.

|  | Acc | Sn | Sp | F-Score | ROC |
|---|---|---|---|---|---|
| Hussain et al[4] | 95.18 | 97.00 | - | 0.9400 | - |
| Proposed Study | 96.05 | 96.22 | 95.91 | 0.9622 | 0.9899 |

## VI. CONCLUSION

We proposed a novel feature extraction method and an effective predictive model for discrimination of acid and alkaline enzymes using discreet protein sequence representation methods piSAAC. In the aforesaid method, we have calculated each terminus and weighted average accuracy of each terminus respectively. Several classification algorithms were employed namely KNN, MLP, KNN, SMO, SVM, Ada boost and DNN. DNN is shown to be the best classifier among the other yielding best result for discriminating enzymes. DNN returned outstanding results over piSAAC of Accuracy 98.01%, the sensitivity of 97.12%, a specificity of 95.87%, F Measure 0.9812 and AUC 0.9500 respectively on dataset1. Our proposed model, shows also the outstanding prediction success rate of accuracy 97.85%, the sensitivity of 97.54%, the specificity of 96.24%, F Measure 0.9774 and AUC 0.9803 respectively on dataset2. Based on these concrete simulation results, it is foreseeing, that the proposed model might be a very powerful and handy tool in the field research studies of enzymes and other proteomics studying fields.